\documentstyle[aps]{revtex}
\input{epsf.tex}
\begin{document}
\draft
\title{Hydrodynamic modes of a 1D trapped Bose gas}
%\title{Hydrodynamic modes in dense trapped ultracold gases}
\author{J. N. Fuchs}
\address{Laboratoire Kastler Brossel,
 Ecole Normale Sup\'erieure*,
24 rue Lhomond, 75231 Paris Cedex 05, France}
\author{X. Leyronas and R. Combescot }
\address{Laboratoire de Physique Statistique,
 Ecole Normale Sup\'erieure*,
24 rue Lhomond, 75231 Paris Cedex 05, France}
\date{Received \today}
\maketitle

\begin{abstract}
We consider two regimes where a trapped Bose gas behaves as a one-dimensional 
system. In the first one the Bose gas is microscopically described by 3D 
mean field theory, but the trap is so elongated that it behaves as a 1D gas
with respect to low frequency collective modes. In the second regime we 
assume that the 1D gas is truly 1D and that it is properly described
by the Lieb-Liniger model. In both regimes we find the frequency of
the lowest compressional mode by solving the hydrodynamic equations.
This is done by making use of a method which allows to find analytical 
or quasi-analytical solutions of these equations for a large class of models
approaching very closely the actual equation of state of the Bose gas.
We find an excellent agreement with the recent results of
Menotti and Stringari obtained from a sum rule approach.
\end{abstract}

%\begin{multicols}{2}

\section{INTRODUCTION}

Among the many experimental tools used to study Bose condensates of 
trapped cold atoms \cite{dgps}, the study of collective modes has 
played a quite important role. Indeed they are of high importance both  
experimentally and theoretically. On the experimental side they provide 
direct in situ informations on the system, which are free of the 
quantitative interpretation of expansion experiments (this is of particular 
importance for dense systems where the standard mean field 
approximation does not work). On the theoretical side the low energy 
collective modes are the elementary excitations and as such they play an 
essential role in the physical understanding of these systems. This is 
clear from the fact that they reduce to phonons for homogeneous 
systems.

The recent developments in this field have shown, among others, a 
growing interest in two different directions. One of them is toward the 
study of systems with reduced dimensionality \cite{angket}. Indeed it is 
experimentally possible to produce anisotropic trapping potentials 
which are strong enough in some directions to freeze the corresponding 
degrees of freedom. In these cases the ultracold atoms are in the ground 
state corresponding to the motion in these directions, since they have 
not enough energy to reach the related excited states. This produces 
systems which are effectively one-dimensional (1D) or two-dimensional 
(2D). These systems are of high fundamental interest since their physics 
is expected to have quite specific properties, different from those of 3D 
systems, which are in particular quite relevant for condensed-matter 
physics. By comparison atomic gases offer particularly clean, 
controllable and simple situations, quite close to model systems.

Another trend is toward the study of strongly interacting systems 
\cite{angket}. 
Indeed most of the experiments have been performed with gases where 
the interactions are weak enough to be properly described by mean field 
theory. We define strongly interacting systems as those for which this 
description is no longer valid. In 3D this corresponds to dense gases. 
For Bose gases the coupling to molecular states has been observed and 
studied very recently in these dense systems \cite{donley}, and one can 
hope to obtain in this way molecular condensates. Another interest is 
to bridge the gap between dilute Bose condensates on one hand, and liquid 
superfluid $ ^{4}$He on the other hand where the interactions are very 
strong. For Fermi gases it is also quite interesting to go to dense 
systems because the BCS-like transition, which is presently very 
actively looked for, is expected to have a much higher critical 
temperature in this regime.

Very recently Menotti and Stringari \cite{Menotti} (MS) have dealt with 
the problem of the collective oscillations in a 1D Bose gas at zero 
temperature, trapped in a very elongated harmonic potential well. They 
considered both the case of a weakly and of a strongly interacting Bose 
gas. The first range covers the high to intermediate density regime, 
where mean field theory is always valid at a microscopic level and the 
system is described by the Gross-Pitaevskii equation \cite{GP}. 
The high density 
limit corresponds to the case where the Thomas-Fermi approximation of 
this equation is valid, and the elongated gas has physically the shape of 
a '3D cigar'. The intermediate density regime corresponds to the case 
where, for the transverse directions, all the particles are in the gaussian 
wavefunction describing the ground state of the transverse harmonic 
potential and only longitudinal degrees of freedom are left. This specific 
situation is called the '1D mean field' regime. For all this range the 
system behaves for the low frequency modes as a 1D system, since the 
transverse collective degrees of freedom correspond to much higher 
frequencies. However for this effective 1D behaviour the system is no 
longer effectively mean field (except in the intermediate density regime), 
i.e. specifically the chemical potential is no longer linear in 1D density 
$n_1$, although the system is still microscopically described by 3D mean 
field. This is because one has to average over the transverse directions, 
which MS did from the Gross-Pitaevskii equation to obtain the chemical 
potential $ \mu (n_1)$ as a function of $n_1$.

In the second range studied by Menotti and Stringari \cite{Menotti}, the 
gas goes from intermediate density (where the interaction is still weak 
and the 1D mean field regime applies) to low density where the 
interaction is effectively strong. For all this range the gas is physically 
in a 1D situation, with 1D particle density $n_1$. A famous example of a 
1D strongly interacting system is 
the Lieb-Liniger \cite{Lieb} model of hard core bosons, of mass $m$, 
interacting via a repulsive delta potential $g_1 \delta (z)$, 
which they solved exactly. The 
paradoxical situation of a gas strongly interacting in a dilute limit (the 
so-called Tonks-Girardeau limit \cite{TG}) 
is actually due to the behaviour of the 
kinetic energy. A characteristic kinetic energy is $\hbar^{2}k^2 /m$, where 
$k$ is typically related to the interparticle distance $d$ by $ k \sim 1/d = 
n_1$. This is small in the dilute regime compared to a typical interaction 
energy $n_1 g_1$, the ratio between these two energies being 
essentially the Lieb-Liniger coupling constant $ \gamma = m g_1 / 
\hbar^{2} n_1$. For a D dimensional space the ratio behaves as $ n_D 
^{1-2/D}$, so it does not depend on density for $D=2$ and decrease 
with density for $D=3$.
Specifically MS studied the Lieb-Liniger model with $ \mu (n_1)$ 
obtained numerically from the Lieb-Liniger solution. In both ranges 
Menotti and Stringari used a sum rule approach to obtain the lowest
frequency mode as a function of frequency. For the three limiting cases
of the 3D cigar, the 1D mean field and the Tonks-Girardeau limit,
they recovered the results which they obtained from the analytical
solution of the hydrodynamic equations.

In the present paper we will use everywhere the hydrodynamic equations
as a starting point. This is a quite natural and general approach
since hydrodynamics is expected to be generically valid to describe the 
dynamics of the system for low frequencies and long wavelengths. In 
particular this approach is quite natural for dilute Bose condensates 
since hydrodynamics appear as a consequence of the Gross-Pitaevskii 
equation. Actually this is quite clear in 3D, but this link is more 
ambiguous in 1D. Anyway we will assume that in 1D hydrodynamics is 
a valid starting point to obtain the low frequency collective modes of the 
system. 

Quite recently two of us \cite{rcxl} (CL) considered how the 
fruitful hydrodynamic approach for collective oscillations in trapped 
Bose condensates, developped in the mean field regime \cite{str}, could 
be extended for strongly interacting systems where mean field is no 
longer valid. It was shown that, in this general case, the linear 
hydrodynamics could still be written in a quite convenient way. This 
made possible to find a number of specific functional dependence for $ 
\mu (n)$ for which an analytical or quasi-analytical solution could be 
obtained. Conversely in the general case it appears possible to 
approximate $ \mu (n)$ closely enough by some of these specific cases, 
considered as models, to obtain a very good approximation for the 
actual result, both for the frequency and the shape of the modes. This 
method can be applied to any mode. The flexibility of this modeling 
allows even in fact to invert experimental data covering a range of 
density to obtain the corresponding chemical potential $ \mu (n)$. Since 
this method is equivalent to solve in a simple and efficient way the 
hydrodynamic equations, it is the purpose of the present paper to apply 
it to the case of the 1D Bose gas, in order to obtain the mode frequency 
from the solution of the hydrodynamic equations, rather than from the 
sum rule method that MS used.

The paper is organized as follows. In the following section we will 
recall the CL method. Actually their original paper considered only 
explicitely the isotropic 3D case. So we will write the generalization to 
any dimension D, since we are interested in 1D, and present the models 
we will use. Then in the next section we will consider the first case 
investigated by MS where the gas is microscopically 3D mean field, but 
behaves as a 1D system for the low frequency modes. We will show 
that, even if mean field does not apply at a microscopic level, the same 
reduction to an effective 1D problem can be obtained, by generalizing 
the procedure used by Stringari \cite{stringanis} in the case of mean 
field. We will obtain explicitely the 1D effective chemical potential from 
the 3D $ \mu (n)$.  Then, before turning to the case of the 1D Bose 
gas, we will reinvestigate the case of the 3D mean field Fermi gas in an 
isotropic trap, which CL had already considered. This has been 
motivated by our preliminary results on the 1D Bose gas which, 
although already quite reasonable, were not as good as expected. This 
led us to improve our method, essentially by making a first order 
correction to the result to take into account the small difference between 
the model $ \mu (n)$ and its actual value. In order to assess the results 
we have considered the 3D mean field Fermi gas, for which we had 
direct results from the numerical integration of the hydrodynamic 
equations. We have also checked the sum rule method in this case. The 
final result of all this is that we obtain the mode frequency with a 
relative precision of at least $10 ^{-3}$ and often much better. 
Naturally this precision, which was not looked for, is much better than 
necessary to compare with experiment, and it is likely that there will be 
some time before it proves useful to be so precise. On the other hand 
there is no reason to put aside this precision since we have it fairly 
easily. Moreover it gives us a very high confidence in our results, all 
the more because we can use different method which agree within this 
precision. Finally we will turn to the 1D Bose gas, present our results 
and compare them with those of MS. The excellent agreement we find 
is a very good check for these two quite different procedures.

\section{SOLVABLE  MODELS  FOR  HYDRODYNAMICS}

The CL approach \cite{rcxl} can be extended quite generally to 
anisotropic traps. We will not dwell here on this general situation since 
we do not intend to make a specific use of it. We are only interested in 
very anisotropic traps. The intermediate situation of moderately 
anisotropic traps is indeed not very convenient theoretically to extract 
informations on the system. Experimentally it is also not so frequently 
used and one rather deals much more often with very elongated or very 
flat traps, which are quasi 1-dimensional or quasi 2-dimensional. Let us 
just mention that, in the $ \alpha - p$ modeling (see below), we are 
restricted to $ \alpha = 2$ in this general situation. On the other hand it 
is possible to extend the use of quasipolynomial models (see below) to 
general anisotropic traps. We will rather show explicitely how the CL 
procedure can be used in any dimension D, having naturally in mind the 
interesting cases D=1 and D=2. This will at the same time allow us to 
provide a short recall of this approach.

As MS and CL we restrict ourselves to the case of the reactive 
hydrodynamics where dissipation is negligible and thermal effects can 
be omitted. This is valid for example at low enough temperature. In this 
situation we have just to write Euler equation $m \, d {\bf  v}/dt = - {\bf  
\nabla} \tilde{\mu } ({\bf  r},t) $ together with particle conservation $ 
\partial n / \partial t + {\bf  \nabla}. (n {\bf  v}) = 0 $ for density $ n 
({\bf  r},t)$. The global chemical potential $ \tilde{\mu } ({\bf  r},t) =  
\mu (n ({\bf  r},t)) + V({\bf  r}) $ has a contribution from the trapping 
potential $ V({\bf  r})$ and a contribution $ \mu (n ({\bf  r},t))$ from 
the fluid itself, where $\mu (n )$ is the equilibrium dependence of the 
chemical potential on the density, as it results from thermodynamics. At 
equilibrium the particle density $ n_{0}({\bf  r}) $ satisfies $ \mu 
(n_{0}({\bf  r})) + V({\bf  r}) = \tilde{\mu }$ where  $\tilde{\mu }$ is 
the constant value over the system of the global chemical potential. In 
particular $\tilde{\mu }$ is equal to the value of the trapping potential at 
the surface of the cloud (we take $ \mu (n=0) = 0$) . For small 
fluctuations we introduce the departure of the chemical potential from its 
equilibrium value $ w({\bf r},t) = \tilde{\mu } ({\bf  r},t) - \tilde{\mu 
}= \mu (n ({\bf  r},t)) -  \mu (n_{0}({\bf  r})) = (\partial \mu / \partial 
n_{0}) \delta n ({\bf  r},t)$ where $ \delta n ({\bf  r},t) = n ({\bf  r},t) - 
n_{0}(r) $ is the density fluctuation. Assuming that the small 
fluctuations occur at frequency $ \omega $ gives $ i \omega  \delta n =  
{\bf  \nabla}. (n_{0} {\bf  v}) =  n_{0}{\bf  \nabla}.{\bf  v} +  {\bf  
v}.{\bf  \nabla} n_{0}$ which, together with Euler equation $ i \omega 
m {\bf  v} = {\bf  \nabla}w $, leads to:
\begin{eqnarray}
n_{0}{\bf  \nabla} ^{2} w+ {\bf  \nabla} n_{0}.{\bf  \nabla}w + m 
\omega ^{2}(\partial n_{0} / \partial \mu) \, w = 0
\label{eq1}
\end{eqnarray}

Let us now restrict ourselves to an isotropic trap $ V(r) $ in D 
dimensions, where $r$ is the distance from the origin. The equilibrium 
relation $ \mu (n_{0}(r)) + V(r) = \tilde{\mu }$ gives $ (\partial \mu / 
\partial n_{0})(\partial n_{0} / \partial r) = - V'(r)$ where $V'(r)$
is the derivative of $V(r)$ with respect to $r$. Then for a mode 
with spherical symmetry $ w({\bf r}) = w(r)$ Eq.(1) becomes:
\begin{eqnarray}
rw'' + [ D-1 + r L'(r)] w' - \frac{ m \omega ^{2} r }{ V' (r) }L'(r) w 
= 0
\label{eq2}
\end{eqnarray}
where we have defined $ L(r) = \ln (n_{0}(r))$ with $ L'(r) = dL/dr $. 
More generally a mode with an angular dependence $ w({\bf r}) = Y 
_{lm} ( \theta , \varphi ) w(r) $ in 3D leads to:
\begin{eqnarray}
rw'' + [D-1 + r L'(r)] w' - [ \frac{g(l)}{r} + \frac{ m \omega ^{2} r }{ 
V' (r) }L'(r) ] w = 0
\label{eq3}
\end{eqnarray}
where $ g(l) = l(l+1) $. Similarly in 2D an angular dependence $ w({\bf 
r}) = \exp(\pm il \varphi ) w(r)$ gives the same equation with $ g(l) = l 
^{2} $. Finally in 1D one gets again the same equation with $ g(l) = 0 
$. In this last case it is better to rename the variable $z$ since it is the 
abscissa and goes from $-R$ to $R$, where $R$ is the cloud radius. It 
is then convenient to make explicit the dominant dependence at small 
$r$ by setting $ w(r) = r ^{l} v(r) $ which leads to:
\begin{eqnarray}
rv'' + [ 2l+D-1 + r L'(r)] v' - [\frac{ m \omega ^{2} r }{ V' (r) } - l ] 
L'(r) v = 0
\label{eq4}
\end{eqnarray}
This equation is valid in 3D and in 2D. It is also true in 1D with $l$ 
taking only two values : $l=0$ corresponding to the modes $ w(z)$ 
even with respect to $z$, or $l=1$ corresponding to the modes odd 
with respect to $z$.

We focus now on the most common case of the harmonic trap $ V(r) = 
\frac{1}{2} m \Omega ^{2} r^{2}$ where Eq.(4) reduces to:
\begin{eqnarray}
rv'' + [ 2l+D-1 + r L'(r)] v' - (\nu ^{2} - l ) L'(r) v = 0
\label{eq5}
\end{eqnarray}
with $ \nu ^{2} =  \omega ^{2} /  \Omega ^{2}$. This equation has the 
quite convenient property to be scale invariant, if the same change of 
scale is naturally made for $L(r)$. In particular it is unchanged under 
the replacement $ r \rightarrow r/R $ so we can take the cloud radius 
$R$ as unity in the following. More generally the change of variable $ y 
= r ^{ \alpha }$ leads only to a modification of the constants in Eq.(5), 
provided again that the same change is made for $L(r)$. One finds 
explicitely:
\begin{eqnarray}
y \frac{d ^{2}v}{ dy ^{2}}  + (\Delta + y  \frac{dL}{ dy})  
\frac{dv}{ dy} - \frac{\nu ^{2} - l }{\alpha } \frac{dL}{ dy} v = 0
\label{eq6}
\end{eqnarray}
where $ \Delta = 1+ \frac{2l+ D-2}{\alpha }$. Finally we note that 
$L'$ does not change if the density $ n(r)$ is multiplied by a constant. 
So the absolute scale in density disappears and we have to deal only 
with the reduced density $ \bar{n}(r) \equiv n(r)/n(0)$. Similarly we 
introduce the normalized local chemical potential $\bar{\mu }(r) \equiv 
\mu (n(r))/\mu (n(0))$ where $ \mu (n(0))$ is related to the gas radius 
$R$ by $ \mu (n(0)) = \frac{1}{2} m \Omega ^{2} R^{2}$. This leads 
to $ \bar{\mu } = 1 - r^{2}$, when $r$ is expressed in units of $R$.

When one takes the model $ dL/dy = - p / (1-y) $, Eq.(6) 
reduces to the hypergeometric differential equation:
\begin{eqnarray}
y (1-y) \frac{d ^{2}v}{ dy ^{2}} + [\Delta - y (p+\Delta)] \frac{dv}{ 
dy}+ p \frac{\nu ^{2}-l }{\alpha } v = 0
\label{eqhyper}
\end{eqnarray}
The general solution of this equation, regular for $y = 0$, is the 
hypergeometric function $ F(a,b; \Delta;y)$, with $a$ and $b$ defined 
by $ a + b = p + \Delta - 1 $ and $ ab = - p \frac{\nu ^{2} - l }{\alpha 
}$. The boundary condition \cite{rcxl} at the surface of the cloud 
$y=1$ give the further condition $a = - n$ where $n$ is a non negative 
integer. In this case the solution is just a polynomial.
This leads for the normal mode frequencies to the explicit result:
\begin{eqnarray}
\frac{\omega ^{2}}{\Omega ^{2}} = l + \frac{\alpha }{p} \: n \: ( n + 
p + \Delta - 1)
\label{eqfreq}
\end{eqnarray}
In particular for the lowest frequency mode which we will consider 
below, corresponding to $l=0$ and $n=1$, the solution is merely $ v = 
(p+ \Delta ) y - \Delta $, for frequency $ \nu ^{2}= \alpha ( 1+ \Delta 
/p)$. The above model, which we call the $ \alpha -p$ model in the 
following, corresponds explicitely to the equilibrium density 
$\bar{n}(r)= (1-r^ \alpha )^p$, which arises from the equation $  
\bar{\mu } = 1 - (1-\bar{n}^{1/p}) ^{2/ \alpha }$ linking chemical
potential to density. In particular for $ 
\alpha =2$, this is merely the power law $  \bar{\mu } = 
\bar{n}^{1/p}$. Actually, if we consider 1D bosons, the 3D cigar, the 
1D mean field and the Tonks-Girardeau limit satisfy precisely this 
functional dependence \cite{Menotti} $ \bar{\mu } = \bar{n} ^{1/p}$ 
between chemical potential and density, with the respective values $p=2 
$, $p=1$ and $p=1/2$. This leads to the following result for the even 
mode frequencies ($l=0$):
\begin{eqnarray}
\frac{\omega ^{2}}{\Omega ^{2}} = 2 n + \frac{n}{p} ( 2 n - 1)
\label{eqfreqeven}
\end{eqnarray}
and:
\begin{eqnarray}
\frac{\omega ^{2}}{\Omega ^{2}} = (1 + \frac{n}{p}) ( 2 n + 1)
\label{eqfreqodd}
\end{eqnarray}
for the odd mode frequencies ($l=1$). In this last case $n=0$ gives the 
'dipole mode' (oscillation of the gas as a whole) with frequency $ 
\omega = \Omega $. These results are in agreement with MS (their 
result gives the even or odd frequency modes, depending on the parity 
of their integer $k$). 

In addition to the simple $ \alpha -p$ model considered above, CL 
found also for $v(y)$ a wide class of quasi-polynomial solutions for more 
complex models. These quasi-polynomials are actually very rapidly 
converging series, which in pratice behave as polynomials (one can 
safely cut them off above some order) because the variable $y$ is 
restricted by $ 0 \le y \le 1$. The corresponding models are 
$ dL/dy = - \sum _{k=0} ^{K}p _{k}y ^{k} / (1-y) $ and can be 
considered as series expansion for $ (1-y) dL/dy$ around $y=0$ with 
increased accuracy. They have in general $K+1$ parameters in addition
to $ \alpha $. In the 
following we will only use the $K=1$ model ($K=0$ is the $ \alpha -
p$ model) which has three parameters $ \alpha , p_0$ and $p_1$ (we 
will call it the 3 parameters quasi-polynomial model). It is solution of:
\begin{eqnarray}
y (1-y) \frac{d ^{2}v}{ dy ^{2}}  +  (q _{2}y ^{2}+ q _{1}y + q 
_{0}) \frac{dv}{ dy} + (r _{1}y + r _{0}) v = 0
\label{equasipol}
\end{eqnarray}
with $ q _{2} = - p _{1}, q _{1} = -(\Delta + p _{0}), q _{0} = 
\Delta, r _{1} = p _{1} \frac{\nu ^{2} - l }{\alpha }$ and $r _{0} = p 
_{0} \frac{\nu ^{2} - l }{\alpha }$. The solution $ v =  \sum _{n=0} 
^{N }a _{n}y ^{n}$ is found by solving the recursion relation 
\cite{rcxl}  between the $ a _{n}$'s resulting from 
Eq.(\ref{equasipol}). Furthermore one requires $ a _{N+1} = 0 $ 
which provides the equation allowing to find the mode frequency $\nu 
^{2}$. The cut-off order $N$ is chosen large enough to insure perfect 
convergence. In practice we have taken in our calculation $N=9$ or 
$10$, which is large enough for excellent convergence and small 
enough for very easy numerical calculations.

\section{REDUCTION  FROM  A  3D  TO  AN  EFFECTIVE  1D 
PROBLEM}

As we already discussed in the introduction there is one clear limit 
where a gas cloud can be considered as a one-dimensional system. This 
is the case where the transverse trapping potential is so strong, 
compared to temperature or interaction, that only degrees of freedom 
corresponding to motion along the weak longitudinal trapping direction 
are left. The transverse degrees of freedom are completely frozen, the 
atoms being all in the ground state with respect to transverse motion. 
However there are weaker conditions under which, with respect to the 
modes, one has still an effective one-dimensional physics. Indeed if we 
deal with low enough frequencies, the transverse degrees of freedom 
will not be excited and we still have to deal only with the longitudinal 
degrees of freedom, which corresponds to a one-dimensional physical 
situation. This occurs for very elongated cigar-shaped traps where the 
transverse trapping is not too strong. This corresponds to the 3D cigar 
regime studied by Menotti and Stringari \cite{Menotti}. In their study 
MS used mean field theory to obtain the effective relation between 
chemical potential $ \mu $ and one-dimensional density $n_1$ to be 
used in their 1D treatment. Here we want to show how this result can 
be extended beyond mean field, and how in the framework of our 
approach we can derive the one-dimensional equation for the modes in 
the 3D cigar regime. We basically follow the procedure of Stringari 
\cite{stringanis} which amounts to integrate Eq.(1) over the transverse 
variables.

We consider an harmonic cigar shaped anisotropic potential, very 
elongated along the $z$ axis:
\begin{eqnarray}
V(r_{\perp},z) = \frac{1}{2} m ( \omega _{\perp}^{2} r_{\perp} 
^{2} + \omega _{z}^{2}z ^{2})
\label{eq5b}
\end{eqnarray}
with $ \omega _{z} \ll \omega _{\perp}$. 
The gas has a transverse Thomas-Fermi radius $ R(z)$ which depends 
on the location on the $z$ axis. Since on the border of the cloud the 
local chemical potential $ \mu (n_{0}({\bf  r}))$ is zero, the 
equilibrium relation between potential and chemical potential gives:
\begin{eqnarray}
\frac{1}{2} m ( \omega _{\perp}^{2} R ^{2} (z) + \omega _{z}^{2}z 
^{2}) = \frac{1}{2} m \omega _{z}^{2}Z ^{2}
\label{eq6b}
\end{eqnarray}
where $Z$ is the maximal extension of the cloud along the $z$ axis. We 
integrate Eq.(1) over the transverse position $ {\bf  r}_{\perp} $, at 
fixed $z$, inside the disk $ r _{\perp} \leq R(z)$. The two first terms of 
Eq.(1) are $ {\bf  \nabla} (n_{0}.{\bf  \nabla}w) = { \nabla} _{\perp} 
(n_{0}.{ \nabla}_{\perp}w )+ { \nabla} _{z} (n_{0}.{ \nabla}_{z}w 
)$. When the first term is integrated, from the divergence theorem it 
gives $ 2 \pi n_{0}.{ \nabla}_{\perp}w$ to be evaluated for $ r 
_{\perp} = R(z) $. This is zero since the density $ n_{0}$ is zero at the 
cloud surface.

Now comes the fact that the transverse degrees of freedom are not 
excited. In this case we have locally equilibrium in the transverse 
direction. This implies that the chemical potential fluctuation $ w( {\bf  
r})$ does not depend on $ {\bf  r}_{\perp}$ and depends only on $z$, 
so we have only to deal with $w(z)$. Hence $ {\nabla}_{\perp}w = 0$ 
(at lowest order in $(\omega _{z}/\omega _{\perp}) ^{2}$). Then in 
order to integrate $ { \nabla} _{z} (n_{0}.{ \nabla}_{z}w )$ over the 
transverse variable, it is more convenient to use, instead of 
$r_{\perp}$, the local equilibrium chemical potential $ \mu (n_{0}({\bf  
r}))$ as a variable. From the equilibrium condition $ \mu (n_{0}({\bf  
r})) + V({\bf  r}) = \tilde{\mu }$ we have $ d \mu = - m  \omega 
_{\perp}^{2} r_{\perp} dr_{\perp} $ and Eq.(1) leads us to:
\begin{eqnarray}
[\: \int_{0}^{ \mu (z)} \! \! d \mu  \:  n_{0}] { \nabla}^{2}_{z}w + 
{ \nabla} _{z} [\: \int_{0}^{ \mu (z)} \! \! d \mu  \:  n_{0}].{ 
\nabla}_{z}w
+ m \omega ^{2} w  \: \int_{0}^{ \mu (z)}\! \! d \mu \frac{ \partial 
n_{0}}{ \partial \mu } = 0
\label{eq7b}
\end{eqnarray}
Here the integration over $ \mu $ goes from zero (corresponding to the 
cloud border) to $\mu (0,0,z) \equiv  \mu (z) \equiv 
\frac{1}{2} m  \omega _{z}^{2}( Z 
^{2} - z ^{2})$ which is the local chemical potential on the $z$ axis. 
Since $w$ depends only on $z$ we have taken it out of the integral, 
together with its derivatives. We are left with $ \int  \! d \mu  \: \partial 
n_{0} / \partial \mu = n_{0}(\mu (z))$ where $ n_{0}(\mu (z))$ is the 
equilibrium density on the $z$ axis. Let us introduce:
\begin{eqnarray}
L(z) = \ln [\int_{0}^{ \mu (z)} \! \! d \mu  \:   n_{0}(\mu) ]
\label{eq8}
\end{eqnarray}
The argument of the logarithm is just the transverse average $ \int  \! d 
\mu  \:     n_{0}(\mu) \sim \int  \! d {\bf  r}_{\perp} \:  n_{0}({\bf  r}) 
$ of the three-dimensional density. We have $ { \nabla}_{z} \exp L(z) 
= n_{0}(\mu (z)) d \mu (z) / dz$ with $ d \mu (z) / dz = - m   \omega 
_{z}^{2}z  \equiv - dV(z) / dz $ where $ V(z) = \frac{1}{2} m  \omega 
_{z}^{2}z ^{2} = V(0,z) $ is just the trapping potential on the $z$ 
axis. Actually for our derivation we do not need an harmonic 
dependence on $z$ for the trapping potential. On the other hand it is 
necessary to have an harmonic dependence on  $ {\bf  r}_{\perp} $, for 
our change of variables. Taking all this into account Eq.(14) becomes:
\begin{eqnarray}
w'' + L'(z) w' - \frac{ m \omega ^{2} }{ V' (z) }L'(z) w = 0
\label{eq9b}
\end{eqnarray}
This is exactly the equation we have already obtained for 1D situations. 
In these cases we would have defined $ L(z) = \ln n_{0}(z)$. We see 
that we have just to replace the one-dimensional equilibrium density $ 
n_{0}(z)$ by the transverse average $ n_{1}(z) 
\equiv \int  \! d {\bf  r}_{\perp} \:  n_{0}({\bf  r}) 
= 2 \pi \int  \! d \mu  \:  n_{0}(\mu) / m \omega_{\perp} ^{2}$
of the 
three-dimensional density. This sounds a physically very reasonable 
result. In the mean field case we have $ \mu = g n_{0}$ which leads to 
$ \int  \! d \mu  \:    n_{0}(\mu) = \mu ^{2}(z)/2g $ and 
Eq.(16) becomes explicitely for a harmonic potential:
\begin{eqnarray}
\frac{1}{4} ( Z ^{2} - z ^{2}) w'' - z w' + \nu ^{2}w = 0
\label{eq10}
\end{eqnarray}
with $ \nu ^{2} = \omega ^{2}/ \omega _{z}^{2}$. This is as 
expected in agreement with Stringari \cite{stringanis}. The above mean 
field relation $ n_{1}(z) = \pi \mu ^{2}(z)/m \omega_{\perp} ^{2}g $ 
with $ g = 4 \pi \hbar^2 
a/m$ is also in agreement with MS result $ \mu = 2 \hbar  \omega 
_{\perp} (a n_1)^{1/2}$ for the 3D cigar regime. This makes clear that 
our above treatment works only for this 3D cigar regime, since the 
transition to the 1D regime requires to include quantum effects, which 
are beyond the Thomas-Fermi approximation we had to use, 
consistently with hydrodynamics. Note finally that we could obviously 
apply the same treatment to a 2D pancake geometry.

\section{FIRST  ORDER  CORRECTION}

As we have already explained in the introduction, we have improved the 
modeling used by CL and presented above by including a first order 
correction to the straight model evaluation of the mode frequency. This 
correction will be used in the two following sections, but we present it 
independently in this section for clarity. So let us now consider the 
possibility of correcting to first order the small difference between our 
model evaluation of the mode frequency, obtained for the model $ L 
_{0}(r)$ (this can be the $ \alpha -p$ model or the 3 parameters quasi-
polynomial model), and the actual result corresponding to the true $ 
L(r)$. This can be done conveniently for example by converting Eq.(6) 
into a second order differential equation which is formally identical to a 
Schr\"{o}dinger equation (that is without first derivative). This is 
obtained by the change of function $ \psi (y ) = v(y) ( y ^{ \Delta } 
n_{0}(y)) ^{1/2} $ which leads to:
\begin{eqnarray}
-\frac{d ^{2} \psi  }{ dy ^{2}}  + V(y) \psi = 0
\label{eq7t}
\end{eqnarray}
with the effective potential:
\begin{eqnarray}
V(y) = \frac{1}{2} L'' + \frac{1}{4} L^{'2} + (\frac{\nu ^{2} - l 
}{\alpha }+ \frac{ \Delta }{2} ) \frac{ L'}{y}+ \frac{ \Delta ( \Delta -
2)}{4 y ^{2}} 
\label{schrod}
\end{eqnarray}
Eq.(\ref{schrod}) is a Schr\"{o}dinger equation corresponding to zero 
energy (and $ \hbar ^{2}/2m = 1$). If we have a change $ \delta L = L 
- L _{0}$, this gives a corresponding change for $V(y)$ and a variation 
of the energy which can be calculated by conventional first order 
perturbation theory. In order to keep the energy equal to zero, we have 
also to give a compensating variation $ \delta \nu ^{2}$ for the 
frequency, which is just the correction we are looking for. Explicitely 
this gives:
\begin{eqnarray}
\int_{0}^{1} \! dy \: [\frac{1}{2} \delta L'' + \frac{1}{2} L' \delta L' 
+ (\frac{\nu ^{2} - l }{\alpha }+ \frac{ \Delta }{2} ) \frac{ \delta 
L'}{y}+ \frac{\delta  \nu ^{2}}{ \alpha }  \frac{ L'}{y}]  \psi ^{2}= 0
\label{eq9t}
\end{eqnarray}
The $ \delta L''$ term can be integrated by parts (the integrated term is 
zero for the range of parameters we are interested in). When one uses $ 
L'= n' _{0}/ n _{0}$ and writes $ \psi (y)$ in terms of $v(y)$, one 
gets finally:
\begin{eqnarray}
\frac{\delta  \nu ^{2}}{ \alpha } \int_{0}^{1} \! dy \: (-n' _{0}) \, y ^{ 
\Delta -1} v ^{2} =  \int_{0}^{1} \! dy \: \delta L' \, n _{0} \, y ^{ 
\Delta -1} v  [\frac{\nu ^{2} - l }{\alpha } v - y v']
\label{eqcorr}
\end{eqnarray}

\section{THE  3D  MEAN  FIELD  FERMI  GAS}

For the reasons presented in the introduction we study here, on the 3D 
mean field Fermi gas, how to improve the results obtained \cite{rcxl} 
by CL, concentrating specifically on the lowest frequency compressional
mode. The first point to consider is how to best approach the actual $ 
\bar{ \mu }(\bar{n})$ by a model $\bar{ \mu } _{mod}(\bar{n})$. 
This has to be done by minimizing some estimator $I$ of the difference 
between our model and the actual physical equation.
Since we have been interested in precision evaluation, the choice of the 
estimator $I$ of the difference between our model and the actual 
physical equation of state becomes a relevant one. We have considered 
mainly three estimators. The two first ones are the pretty obvious:
\begin{eqnarray}
I ^{2} = \int_{0}^{1} d\bar{n} [ \bar{ \mu }(\bar{n}) - \bar{ \mu } 
_{mod}(\bar{n})] ^{2}
\hspace{2cm} {\rm and} \hspace{2cm}
I ^{2} = \int_{0}^{1} d\bar{ \mu } [\bar{n} (\bar{ \mu }) - \bar{n}  
_{mod} (\bar{ \mu })] ^{2}
\label{eqa}
\end{eqnarray}
and there is no reason why one of them should be systematically better 
than the other one. Note that one way to reconcile these two estimators 
would be to introduce a third one where, instead of dealing with the 
'horizontal' or 'vertical' distance between the two curves, one would 
consider the distance between the two curves perpendicular to them. 
However, since this is somewhat more complicated and does not bring 
a decisive advantage, we have not implemented this solution. On the 
other hand all these estimators do not introduce a specific weighting 
while one could wonder for example if the high density regions are not 
more important than the low density ones. One possible way to 
introduce such a weighting is to require that the result of the first order 
correction we perform is as small as possible. However this correction 
Eq.(\ref{eqcorr}) contains the shape of the mode itself, which is 
naturally not known before we have calculated the estimator and found 
its minimum. A possible way out of this dilemna is to retain only the 
factor in the correction which does not contain the mode and take for 
example:
\begin{eqnarray}
I =  \int_{0}^{1} \! dy \: \, y ^{ \Delta -1} n _{0} \, | \delta L' |
\label{eqb}
\end{eqnarray}
However this estimator proved in our tests to be no better and even 
somewhat worse than the simpler estimators Eq.(\ref{eqa}). In the end 
this question of the estimator choice turned out to be a secondary one 
because, once the first order correction is made, either of the estimators 
Eq.(\ref{eqa}) is good enough and gives excellent precision. In the same 
spirit, once an estimator is chosen the
minimization process may be difficult when there are different minima 
in competition (a well known problem in spin glasses for example). 
However because the corrected results are anyway very good, there is 
no need to go to the precise minimum, and an approximate minimization 
is enough.

For the 3D mean field Fermi gas, with reduced units, the chemical 
potential is given in terms of the density by:
\begin{eqnarray}
\bar{ \mu } = \frac{3  \bar{n} ^{2/3} - 2 \lambda  \bar{n}}{ 3 - 2 
\lambda }
\label{eqaa}
\end{eqnarray}
where $ \lambda $ is the coupling constant which goes from $ \lambda = 0 $ 
for the free Fermi gas to $ \lambda = 1 $ when the attractive interaction 
is strong enough to produce a collapse. It is given by $ \lambda = 2 k_F 
|a| / \pi $ in terms of the negative scattering length $a$ and the 
Fermi wavevector $k_F$. Naturally the interest of this study is that we 
have the results of the direct numerical integration of the hydrodynamic 
differential equation, which gives us a benchmark for our modeling. 
Even this direct integration has a limited precision, which in our case is 
not extremely high because there is no reason to work for very high 
precision. So CL used the classical Runge-Kutta routine with standard 
precision. The resulting numerical noise can be estimated directly from 
inspection of their results for  $ \nu ^{2}$ as a function of $ \lambda $ 
and its first order difference. We have an absolute precision of  $ 10 ^{-
3}$ for these results on $ \nu ^{2}$.

\begin{figure}
\centering
\vbox to 70mm{\hspace{0mm} \epsfysize=7cm \epsfbox{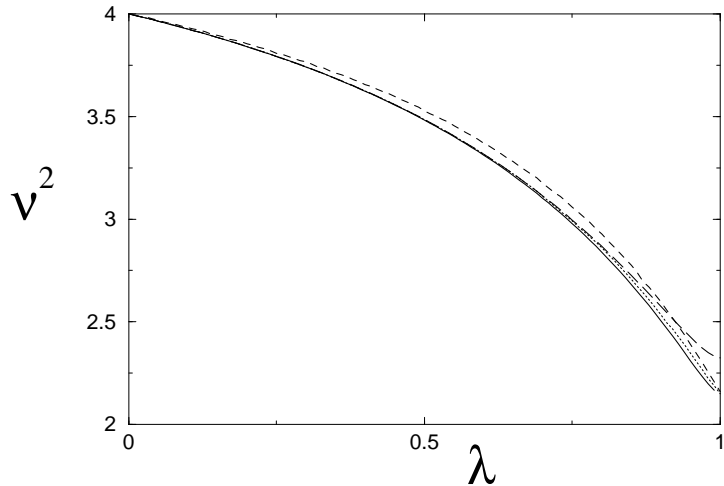} }
\caption{Reduced lowest compressional mode frequency $ \nu ^{2}=
( \omega / \Omega ) 
^{2}$ for a Fermi gas within the mean field approximation as a 
function of the coupling constant $ \lambda $. Full line: exact result 
from the numerical integration of the hydrodynamic equation. Long-
dashed line: sum rule result. Short-dashed line: zeroth order $ \alpha - p 
$ model. Dotted line: zeroth order 3 parameters quasi-polynomial 
model.}
\label{1H}
\end{figure}

\begin{figure}
\centering
\vbox to 70mm{\hspace{0mm} \epsfysize=7cm \epsfbox{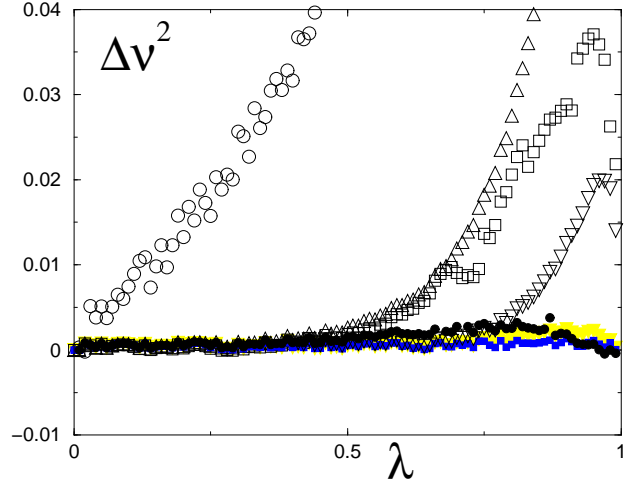} }
\caption{Difference $ \Delta \nu ^2$ of the reduced mode frequency 
between various model results and direct integration of the 
hydrodynamic equation. Open circles: zeroth order $ \alpha - p $ model. 
Open up-triangles: sum rule method. Open squares: 3 
parameters quasi-polynomial model with first estimator in Eq.(22). 
Open down-triangles: 3 parameters quasi-polynomial model with second 
estimator in Eq.(22). Filled circles: corrected $ \alpha - p $ model. 
Filled squares: corrected 3 parameters quasi-polynomial model 
with first estimator in Eq.(22). Filled down-triangles: corrected 3 parameters 
quasi-polynomial model with second estimator in Eq.(22).}
\label{2H}
\end{figure}

The results of our calculations for $ \nu ^2$ as a function of $ \lambda 
$ are summarized in Fig. \ref{1H}. It is clear that all the approximate 
methods give quite reasonable agreement with direct integration. So we 
have plotted in Fig. \ref{2H}, with a much magnified scale, the 
difference between our various approximate calculations and the 
direct integration. As found by CL the zeroth order $ \alpha - p $ model 
gives already quite satisfactory results. Indeed the maximum deviation 
from the direct integration is found around $ \lambda \approx 0.9$. For  
$ \lambda = 0.9$ it gives $ \nu ^{2}= 2.59 $ compared to 2.505  from 
direct integration. It is easy to see why this region for $ \lambda $ is 
more difficult for $ \alpha - p $ modeling. This can be understood from 
the behaviour of the equilibrium density $ n_{0}(r)$ near the origin. 
Since $ \bar{\mu } = 1 - r^{2}$ we have in general $ n_{0}(r) \approx 
n_{0}(0) + {\mathcal O} (r^{2})$. On the other hand at the collapse $ 
\lambda  = 1$ we have $ \partial \mu / \partial n_{0} = 0 $ and near the 
origin  $ n_{0}(r) \approx n_{0}(0) + {\mathcal O} (r)$. Near the 
collapse $ n_{0}(r)$ is, so to speak, switching from one behaviour to 
another. So it is clear that its analytical behaviour will be more complex 
and its modeling will be 
accordingly more difficult.

Making use of the 3 parameters quasi-polynomial model makes a very 
important improvement as can seen in Fig. \ref{1H} and  \ref{2H}. 
Indeed for  $ \lambda \begin{array}{c}<\vspace{-2,5mm} 
\\\sim\end{array} 0.4 $ the results are within the noise of the direct 
integration, that is within  $ 10 ^{-3}$ from the exact result. Then for 
higher $ \lambda $ it starts to deviate with a maximum deviation of 0.03 
for $ \lambda = 0.9 $, then it gets back essentially to the exact value at 
the collapse. 

For comparison we have also plotted on Fig. \ref{1H} and  \ref{2H} 
the result from the sum rule method \cite{str}, which is basically a 
variational method. We have adapted it in the following way to our 
present problem. We start with the expression given by MS for the 
monopole frequency \cite{Menotti} $2 \, \omega_{M}^{-2}=-
d\,\ln\langle r^2\rangle / d\,\Omega^2$ (here we do not use the cloud 
radius as unit length). This relation is more conveniently expressed in 
terms of the oscillator length $l_o=\sqrt{\hbar/m\Omega}$ as:
\begin{eqnarray}
\label{eqnmode}
\nu_{M}^2&=& 8\left(\frac{d\,\ln\langle r^2\rangle}{d\,l_o /l_o 
}\right)^{-1}
\end{eqnarray}
where, as above, $\nu_{M}^2=\omega_{M}^{2}/\Omega^2$ and the 
derivative is taken
at constant number of particles $N$.
Now, we express $\langle r^2\rangle$ as a function of $l_o $, $a$ and 
the
coupling constant $\lambda$ defined above.
In order to calculate $\langle r^2\rangle=\int\,d^{3}{\bf r}\,r^2\,n_0 
(r)$,
we make in this integral the change of variables $t=k_F/k^{*}$ with 
$3\pi^2\,n_0 (r) =k_{F}^3$ and $k^{*}=\pi / 2|a|$. Using the 
equilibrium equation (\ref{eqaa}), we get
$r^2 = l_o ^{4}\,k^{*\,2}\,P(\lambda,t)$, with 
$P(\lambda,t)=\lambda^2-t^2-\frac{2}{3}(\lambda^3-t^3)$.
After a straightforward calculation, we get:
\begin{eqnarray}
\langle\,r^2\rangle&=&\frac{2}{3\pi}\,l_o 
^{10}\,k^{*\,8}\,F_2(\lambda)\\
N&=&\langle 1 \rangle = \frac{2}{3\pi}\,l_o 
^{6}\,k^{*\,6}\,F_0(\lambda)
\end{eqnarray}
where $F_{q}(\lambda)=-\int_{0}^{\lambda}\, dt\,t^3\,P^{(q+1)/2}\, 
\partial P/\partial t$.
From these two equations, we have:
\begin{eqnarray}
\label{eqn1}
d\,\ln\,\langle\,r^2\rangle&=&(F_2^{\prime}/F_2) 
\,d\lambda\,+\,10\,\frac{dl_o}{l_o}\\
\label{eqn2}
d\,\ln\,N&=&(F_0^{\prime}/F_0) \,d\lambda\,+\,6\,\frac{dl_o}{l_o}
\end{eqnarray}
where $F_{q}^{\prime}=dF_{q}/d\lambda$.
Since we take the derivative at constant $N$ in  Eq.(\ref{eqnmode}),
we can eliminate $d\lambda$ from Eq.(\ref{eqn1},\ref{eqn2}) and 
using Eq.(\ref{eqnmode}),
this yields:
\begin{eqnarray}
\nu_{M}^{2}&=&\frac{4}{5-
3\,F_0\,F_{2}^{\prime}/F_{0}^{\prime}F_2}
\end{eqnarray}
The calculation of $\nu_{M}^{2}$ ({\it i.e} the monopole frequency)
therefore simply amounts to calculate four integrals. As it can be seen 
on Fig. \ref{1H} and  \ref{2H}, the sum rule method does remarkably 
well for $ \lambda \begin{array}{c}<\vspace{-2,5mm} 
\\\sim\end{array} 0.4 $ where it gives results within our direct 
integration noise. Then for larger $ \lambda $ it starts to deviate and 
gets worse and worse when one goes to the collapse, where the 
deviation reaches 0.18. This is clearly linked \cite{stringthanks} to the 
fact that the density has a linear dependence on $r$ at the collapse, as 
we have seen, while the sum rule method is best suited when there is a 
quadratic dependence at the center. 
In this respect one has to keep in mind that, even 
near the collapse for the trapped gas cloud as a whole, most of the gas 
has a density far from the one corresponding to the collapse of an 
homogeneous gas, because of the inhomogeneity caused by the 
trapping potential.

When we apply first order perturbative correction, as we have described 
above, we obtain a quite remarkable improvement. This is seen in Fig. 
\ref{2H} (the results have not been plotted in Fig. \ref{1H} since they 
would not be distinguishable from the direct integration ). Indeed the 
results we get from quasipolynomial modeling agree over the whole 
range of values for $ \lambda $ with those of direct integration, within 
the numerical noise of $ 10 ^{-3}$. This is quite gratifying, but 
perhaps not completely unexpected, taking into account that we have 
three adjustable parameters, plus a correction. More surprising is that 
the corrected $ \alpha - p $ model gives also results which are almost 
within numerical noise, though the quasipolynomial results are slightly 
better. Since without correction the quasipolynomial model is clearly 
much better, as it can be seen in Fig. \ref{1H}, this excellent final 
agreement of the $ \alpha - p $ model is perhaps partially coincidental, 
although we find the same feature in the next section for the 1D Bose 
gas.

It is clear that the excellent agreement within $ 10 ^{-3}$ of the results 
from the quasipolynomial and the $ \alpha - p $ model gives us very 
much confidence in the validity of the result itself within this precision. 
We could ignore the direct integration and only infer the final result 
from what we obtain from the various models. This is naturally the 
point of view that we will adopt in the next section when we will study 
1D bosons.

\begin{figure}
\centering
\vbox to 70mm{\hspace{0mm} \epsfysize=7cm \epsfbox{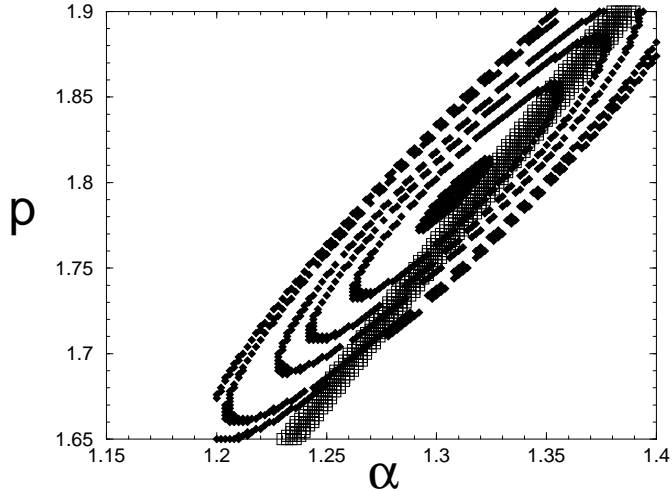} }
\caption{Lines of equal values in the $ \alpha -p$ plane for the first of 
the estimators Eq.(22). Starting from the minimum which is at 3.10$ 
^{-5}$, the black ellipse corresponds essentially to 3.10$ ^{-5}$,the 
dashed lines to 4.10$ ^{-5}$, 5.10$ ^{-5}$,  6.10$ ^{-5}$, 
8.10$ ^{-5}$ and 9.10$ ^{-5}$ respectively for $I ^{2}$. The band 
of open squares corresponds to the region where the reduced frequency 
calculated by the corrected $ \alpha - p $ model is within $\pm$10$ ^{-
3}$ from the direct integration. This band is also shown in Fig. \ref{4H}.}
\label{3H}
\end{figure}

\begin{figure}
\centering
\vbox to 70mm{\hspace{0mm} \epsfysize=7cm \epsfbox{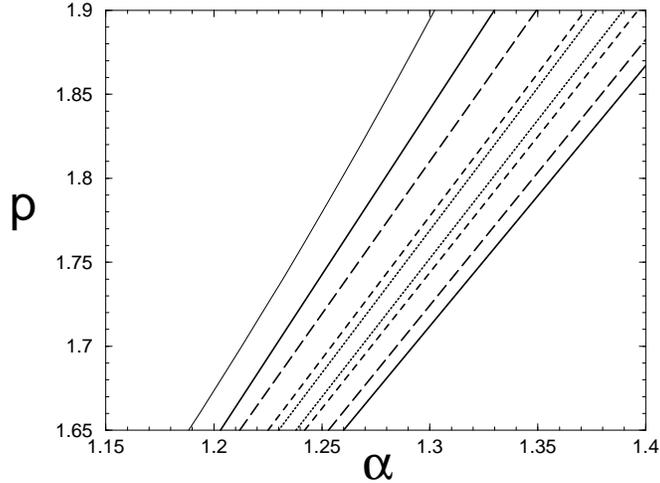} }
\caption{Lines of equal values in the $ \alpha -p$ plane for $ \nu ^{2}$ 
being respectively at $\pm$10$ ^{-3}$(dotted line), $\pm$2.10$ ^{-
3}$(short dashed-line), $\pm$5.10$ ^{-3}$(long dashed-line) and 
$\pm$7.10$ ^{-3}$(full thick line) from the direct integration result $ 
\nu ^2 = 2.505 $. The full thin line corresponds to values in the $ \alpha 
-p$ plane where the first order correction to the $ \alpha -p$ model is 
actually zero.}
\label{4H}
\end{figure}

In this respect, as we mentionned already, it is interesting to note that 
the first order perturbative correction takes essentially care of the 
difference which appears when we use different modeling and/or use 
different estimators and/or use different minima of these estimators. In 
particular, as we have already mentionned, if one requires very precise 
results, it is not always easy to locate the absolute minimum of an 
estimator. The use of the first order perturbative correction solves this 
problem since, as we have seen, the dispersion of the corrected results 
is very small compared to the one of the uncorrected ones. So we are 
allowed a little imprecision in the choice of the parameters of our 
modeling, since the first order perturbative correction will compensate 
for the resulting error in the mode frequency. It is naturally interesting 
to study this point in more details. We have done it for a worst case 
situation $ \lambda = 0.9 $, a value near a maximal dispersion of the 
results as we have seen. We have only considered the $ \alpha - p $ 
model in order to have only a two-dimensional parameter space. We 
have covered the range $ 1.15 \leq \alpha \leq 1.4$ , $ 1.65 \leq p \leq 
1.9$. The lines of equal values for the first of the estimators Eq.(22) are 
drawn in Fig. \ref{3H}. Since we are near the minimum these lines are 
elliptical. One sees that the minimum is quite shallow for the direction in 
the $ \alpha -p$ plane corresponding to the large axis of these ellipses. 
We have also shown in this figure the region where the corrected $ 
\alpha - p $ model gives a result within $\pm$10$ ^{-3}$ from the 
direct integration. It is quite satisfactory that this region goes very near 
the minimum of our estimator (the difference is basically within the 
noise) and is essentially oriented along the large axes of the ellipses 
(naturally we have generically a line in the $ \alpha - p $ plane 
corresponding to any given value of $ \nu ^2$). Again for the corrected 
$ \alpha - p $ model, we have plotted in Fig. \ref{4H} the lines of equal 
values for $ \nu ^{2}$ being respectively at $\pm$10$ ^{-3}$, 
$\pm$2.10$ ^{-3}$, $\pm$5.10$ ^{-3}$ and $\pm$7.10$ ^{-3}$ 
from the direct integration result $ \nu ^2 = 2.505 $. We see that one 
can take values for ($ \alpha , p $) which correspond to an estimator 
three times larger than the minimum and still obtain a result for $ \nu 
^{2}$ which is at worst only within $\pm$5.10$ ^{-3}$ from the exact 
result. The corresponding value of the estimator shows that the average 
deviation between the model and the exact $ \mu (n)$ is roughly $1\%$. 
This shows that, even with a modeling which is not very precise, we 
can obtain a very good result for $ \nu ^{2}$.

When one sees the range of values for the parameters giving essentially 
the correct mode frequency, one may wonder if there is not another 
criterium than the above estimators to find a priori the best set of 
parameters. An interesting possibility in this direction is to say that the 
best parameters should not require any first order correction at all. So 
one should use sets of parameters giving zero correction. These sets 
correspond to a line in the $ \alpha - p $ plane which is shown in Fig. 
\ref{4H}. We have calculated the corresponding mode frequencies. We 
have found that, although the results are fairly independent of the 
parameter set, they fall systematically slightly above, at  2.514 , 
compared to the correct result 2.505 . Actually either of the estimators 
Eq.(\ref{eqa}) turns out to be the best criterium to find the correct result 
(the difference between these two being within the numerical noise). 
This is seen in Fig. \ref{2H} where both estimators have been used.

\section{THE  1D  BOSE  GAS }

We apply now the previously developed method to study the
hydrodynamic modes of the trapped 1D Bose gas with repulsive 
interactions. At
zero temperature, the interactions are characterized by a single
parameter: the 3D $s$-wave scattering length $a$. The
system we consider is the same as the one studied by Menotti and
Stringari \cite{Menotti}. Namely, this is a Bose gas in a very 
anisotropic trap
($\omega_{\perp}\gg \omega_{z}$)
for various one-dimensional configurations. The transverse oscillator 
length
$a_{\perp}=\sqrt{\hbar/m\omega_{\perp}}$ is supposed to be much
larger than the scattering length $a_{\perp} \gg a$. As we mentionned 
already in the introduction, MS identified three limiting regimes 
corresponding physically to decreasing 1D densities $n_1(0)$ at the 
center of the trap. At high density ($n_1 (0)a\gg1$), the system is in the 
radial Thomas-Fermi regime (3D cigar). In this case the chemical 
potential is related \cite{Menotti} to the 1D density by $ \mu = 2 \hbar  
\omega _{\perp} (a n_1)^{1/2}$, as indicated already in section III. 
In the second regime of intermediate densities, it is convenient to
introduce the effective 1D scattering length $a_{1}$ related to the 
Lieb-Liniger coupling constant $g_1$ by $g_1=2\hbar^{2}/ma_{1}$. 
In the limit $a_{\perp} \gg a$, Olshanii \cite{Olshanii} showed that 
$a_{1}=a_{\perp}^{2}/a$  \cite{sign}.
In this intermediate density regime defined by 
$n_1 (0)a_{1}\gg 1 \gg n_1 (0)a$, 
the gas is a 1D quasi-condensate (1D mean-field regime) and the 
chemical potential 
is given by $ \mu = 2 \hbar  \omega _{\perp} a n_1 $ (with our 
conventions $ \mu $ is zero for zero density). 
Finally at low density 
$1\gg n_1 (0)a_{1}$, the gas consists of impenetrable bosons 
(Tonks-Girardeau regime) where $ \mu = \pi ^2 \hbar^2 n_1^2 / 2m$. 

Our needed input, in order to obtain the mode frequencies, is the
equation of state $\mu(n_1)$ for the homogeneous gas at zero
temperature. For the transition between the 3D cigar and the 1D
mean-field regimes (high density domain), it can be obtained from the 
numerical
solution of the 3D Gross-Pitaevskii equation in a cylindrical geometry
\cite{Menotti}. Actually, it appears that the numerical results 
\cite{Chiara} can be very well approximated analytically by:
\begin{equation}
\mu(n_1)=\hbar \omega_{\perp} (\sqrt{1+4n_1 a}-1)
\label{quasi-analytic}
\end{equation}
This formula gives the exact behavior in
the two limiting cases (3D cigar and 1D mean-field) defined above and
interpolates
very well in between (to better than $2\%$). In the following, we
will use this approximate analytic expression which is easier to
manipulate and avoids numerical problems encountered when the 
equation
of state is only known for a fixed set of points. In the high density
domain, we use the dimensionless variable $n_1(0)a$ to go from one regime
to the other.

The equation of state for the transition between the 1D mean-field and 
Tonks-Girardeau
regimes (low density domain) can be obtained from the Lieb-Liniger
solution \cite{Lieb}. Lieb and Liniger gave a closed expression for the 
energy per particle as a
function of the density in the form of an integral equation. It shows
that the chemical potential is a universal
function of $n_1 a_{1}$, which has to be evaluated numerically. 
Menotti and
Stringari \cite{Menotti} calculated this equation of state and made their 
result available \cite{Chiara}. In the following, we use their data. In the 
low density domain, the dimensionless variable we use is $n_1 
(0)a_{1}$.

Using these equations of state,
we compute the mode frequency using the following procedure: 
for each value of $n_1(0)a$ (or $n_1(0)a_{1}$)
we fit the equation of state (for $n_1$ varying between $0$ and $n_1(0)$)
with an analytic model (either $\alpha-p$ or 3 parameters 
quasi-polynomial) using one of the
estimator of equation (\ref{eqa}); the zero order mode frequency is then 
obtained
by inserting the value of the best set of parameters in the formula
giving the mode frequency for the analytic model (Eq.(\ref{eqfreq}) for 
the $ \alpha -p$ model); we then compute the first order perturbation 
correction to the mode frequency.

\begin{figure}
\centering
\vbox to 70mm{\hspace{1mm} \epsfysize=7cm 
\epsfbox{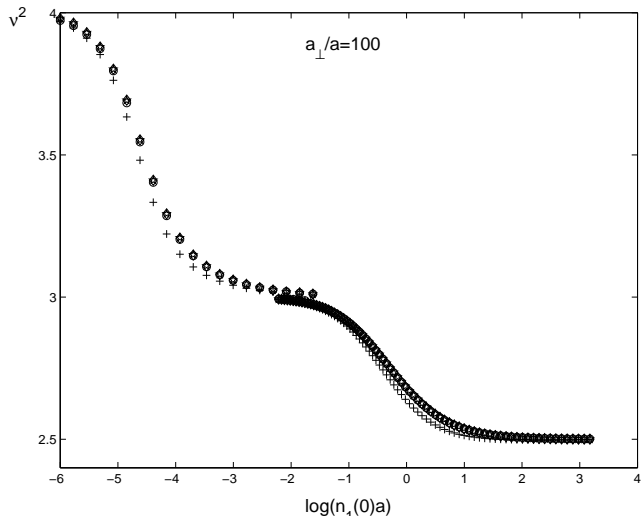} }
\caption{Lowest compressional mode of the trapped 1D Bose gas with
$a_{\perp}/a=100$. The (squared) mode frequency
$\nu^2=\omega^2/\omega_{z}^2$ is plotted as a function of
$\log{(n_{1}(0)a)}$. The crosses ($+$) correspond to the $\alpha-p$
model; the stars ($\star$) to the corrected $\alpha-p$ model;
the circles ($\circ$) to the quasi-polynomial model;
and the diamonds ($\diamond$) to the corrected quasi-polynomial 
model. Actually at this scale all the symbols fall on top of each other, 
except for the crosses.}
\label{figure1b}
\end{figure}

We first discuss the different models used to compute the (squared) 
reduced mode
frequency $\nu^2=\omega^2/\omega_{z}^2$. Fig. \ref{figure1b} 
shows $\nu^2$ as a
function of $n_1 (0)a$ for a system where $a_{\perp}/a=100$. Four 
curves
are plotted. They correspond to the $\alpha-p$ model, the
quasi-polynomial model, and to the same two models corrected to first 
order
of perturbation theory. Actually, the
3 parameters quasi-polynomial model is used with only two
parameters by fixing $\alpha=2$. We checked that, in the case of the 1D
Bose gas and unlike that of the 3D Fermi gas,
it does not make a significant difference
to let the three parameters free or to set $\alpha=2$,
whereas it is much faster and easier to work with only two parameters.

There are much more points plotted between the 3D cigar
and the 1D mean-field regimes (high density domain)
than between the 1D mean-field and the Tonks-Girardeau regime
(low density domain). This is a mere consequence of the fact that we
used an analytic expression for the equation of state in the first
domain Eq.(\ref{quasi-analytic}), whereas in the second one we used
the numerical data of MS. Having a fixed set of
points for the equation of state makes the
numerical evaluation of integrals (needed to compute
the correction $\delta\nu^2$, for example) more difficult because we are 
forced to use a primitive integration algorithm
and, therefore, puts a limit on the precision and on the number of
points that can be calculated safely. The fact that the junction between 
the two domains is not perfect (in the region of the 1D mean-field 
regime) is a consequence of taking a finite value for
$a_{\perp}/a$ instead of the limit $a_{\perp}/a\rightarrow \infty$.
As noted by MS, the 1D mean-field regime can only be identified 
provided
$a_{\perp}/a\gg 1$.

\begin{figure}
\centering
\vbox to 70mm{\hspace{1mm} \epsfysize=7cm 
\epsfbox{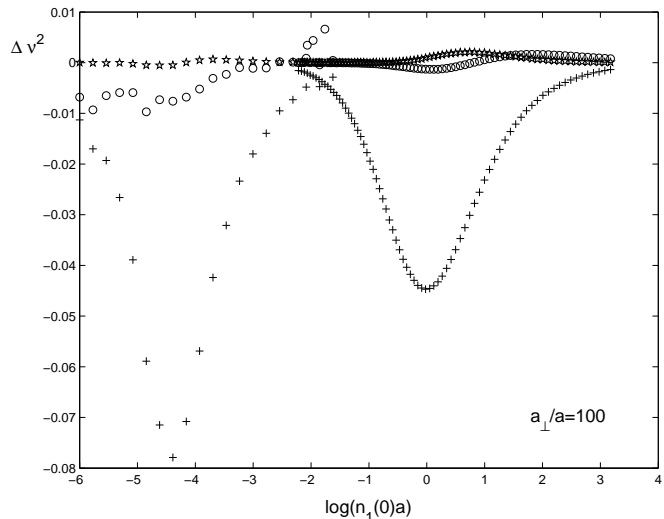} }
\caption{Lowest compressional mode of the trapped 1D Bose gas with
$a_{\perp}/a=100$. The (squared) mode frequency
$\nu^2=\omega^2/\omega_{z}^2$ calculated using the corrected
quasi-polynomial model is taken as a reference. The difference between
this reference and three different calculations ($\alpha-p$, corrected
$\alpha-p$ and quasi-polynomial models) are plotted as a function of
$\log{(n_{1}(0)a)}$. The crosses ($+$) correspond to the $\alpha-p$
model; the stars ($\star$) to the corrected $\alpha-p$ model; and
the circles ($\circ$) to the quasi-polynomial model.}
\label{figure2b}
\end{figure}

In order to compare our various calculations, we look now at our results 
with a much larger scale, which amounts to using a magnifying glass.
We take as reference the results of the corrected quasi-polynomial 
model. As discussed in the preceding section we expect this reference to 
be the best of our results and to be very precise. The difference between 
this reference and the three first calculations
($\alpha-p$, quasi-polynomial and corrected $\alpha-p$ models)
are plotted in Fig. \ref{figure2b}. Two points already made when 
discussing the 3D
Fermi gas are worth stressing again in the case of the 1D Bose
gas. First, once corrected, the results of the $\alpha-p$ and the
quasi-polynomial models agree remarkably well. Second, their 
agreement is at
the absolute $10^{-3}$ level for $\nu^2$.

The $\alpha-p$ model is exact in the three limiting regimes: $\alpha=2$
and $p=2$ in the 3D cigar regime, which gives $\nu^2=5/2$; 
$\alpha=2$
and $p=1$ in the 1D mean-field regime, which gives $\nu^2=3$;
and  $\alpha=2$ and $p=1/2$ in the Tonks-Girardeau regime, which 
gives
$\nu^2=4$. In between these limits, it is apparent on Fig. 
\ref{figure1b} and \ref{figure2b}
that the $\alpha-p$ model has some difficulties in predicting precisely 
the
correct value of the mode frequency. This is what originally motivated
the use of the 3 parameters quasi-polynomial
model and the development of corrections using perturbation theory. 
This difficulty can actually be understood in much the same way as for 
the 3D Fermi gas. For example in the transition between the 3D cigar 
and the 1D mean field, one finds that for large $n_1(0)a$ one has 
essentially $ \mu \sim n_1^{1/2}$ for most of the $n_1$ range, while 
for small $n_1$ this dependence turns into $ \mu \sim n_1$. This 
switch of analytical behaviour is difficult to follow for the simple  
$\alpha-p$ model, hence the somewhat unsatisfactory result.

For the $\alpha-p$ and quasi-polynomial models, using an analytic
expression for the equation of state (high density domain) gives
better results than using numerical data (low density regime), as can
be seen on Fig. \ref{figure2b}. This is linked to the difficulty, 
mentionned above, of using purely numerical data as an entry for $ \mu 
(n_1)$. However, the corrected models agree at the same
level in both domains.

\begin{figure}
\centering
\vbox to 70mm{\hspace{1mm} \epsfysize=7cm 
\epsfbox{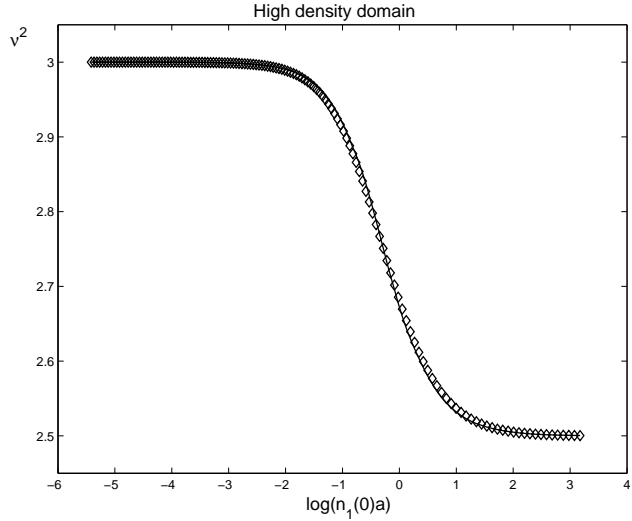} }
\caption{Lowest compressional mode of the trapped 1D Bose gas in the
high density domain. The (squared) mode frequency
$\nu^2=\omega^2/\omega_{z}^2$ is plotted as a function of
$\log{(n_{1}(0)a)}$. The diamonds ($\diamond$) correspond to the
corrected quasi-polynomial model. The full line is the result of
Menotti and Stringari.}
\label{figure3b}
\end{figure}

We will now compare our results with those of Ref.
\cite{Menotti}, first discussing the high density domain.
Menotti and Stringari calculated $\nu^2$ as a function of
$Naa_{\perp}/a_{z}^2$, where
$a_{z}=\sqrt{\hbar/m\omega_{z}}$ is the longitudinal oscillator
length and $N$ the number of particles. In this paper, we calculate
the same quantity as a function of the more convenient $n_1(0)a$, 
which
is an increasing function of $Naa_{\perp}/a_{z}^2$.
Finding the relation between these two numbers amounts to obtaining 
the
normalized density profile. This can be done analytically in the two
limiting regimes \cite{Menotti} and numerically in between. This allows
to plot the MS result for the mode frequency and our corrected
quasi-polynomial result (which we expect to be the most precise of our 
results)
on the same graph, see Fig. \ref{figure3b}. It shows excellent 
agreement, at
the $10^{-2}$ level. Asking for agreement at a better level is not
sensible here, as we used an analytic expression approximating the 
equation
of state within a few percent and not the exact equation of state.
This shows up in the fact that for high densities
($n_1 (0)a>1$), the corrected quasi-polynomial seems to predict values 
of
the mode frequency that are higher than the result of MS.
This would be in contradiction with the fact that the sum rule approach 
they used is known to be an exact upper bound \cite{str}.

As a further check on our method, we also calculated the
expected frequency for two experiments. For this purpose, the exact
equation of state was used, in the form of the numerical solution of
the 3D Gross-Pitaevskii equation \cite{Chiara}. For the experiment
of Ref. \cite{Gorlitz}, where $n_1 (0)a=0.22$, we find 
$\nu^2=2.845\pm0.003$ and
$\nu^2=2.911\pm0.002$ for the experiment of Ref. \cite{Schreck}, 
where
$n_1 (0)a=0.10$. This is in complete agreement with the values first
obtained by MS.

\begin{figure}
\centering
\vbox to 70mm{\hspace{1mm} \epsfysize=7cm 
\epsfbox{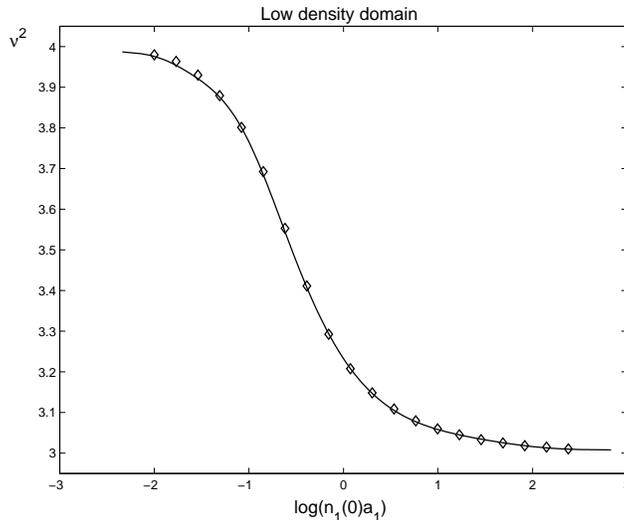} }
\caption{Lowest compressional mode of the trapped 1D Bose gas in the
low density domain. The (squared) mode frequency
$\nu^2=\omega^2/\omega_{z}^2$ is plotted as a function of
$\log{(n_{1}(0)a_{1})}$. The diamonds ($\diamond$) correspond to 
the
corrected quasi-polynomial model. The full line is the result of
Menotti and Stringari.}
\label{figure4b}
\end{figure}

We consider finally the low density domain. As in the high density 
domain,
we first have to relate the dimensionless variable we use $n_1 
(0)a_{1}$
to $Na_{1}^2/a_{z}^2$ used in Ref. \cite{Menotti}. This is done
analytically in the 1D mean-field and in the Tonks-Girardeau regime
\cite{Menotti} and numerically in between. The mode frequency
obtained by the corrected quasi-polynomial model and that calculated by
MS are plotted in Fig. \ref{figure4b}. The agreement is again very 
good.

The results of MS are in very good agreement with our result over the
whole range of $n_1 (0)a$. In the case of the 1D Bose gas, the sum 
rule approach \cite{str}
seems to give not only an upper bound but to be quite near the exact 
result. As already
discussed in the case of 3D Fermi gas, this is related to the fact that
the density has a quadratic dependence on $r$ near the
center of the trap. In other words, the parameter $\alpha$ is always very 
close to $2$.

\section{CONCLUSION}

In this paper we have calculated, as a function of density at the center
of the trap, the frequency of the lowest compressional mode of a 1D 
trapped Bose gas, taking the reactive hydrodynamic equations as a starting
point. We have considered two density regimes. In the first one the density
decreases from high to intermediate, but the Bose gas is always described
microscopically by 3D mean field theory. In the high density region the
Bose gas has the shape of a 3D cigar, while in the intermediate density
region the atoms are in the one-particle ground state with respect to the 
transverse motion and the gas behaves as a 1D system. Nevertheless for all 
this regime the low frequency modes are described by 1D hydrodynamic 
equations, but the effective equation of state Eq.(\ref{quasi-analytic}) 
is no longer given by mean field.
The second regime that we have considered goes from intermediate to low
density and we have taken the Lieb-Liniger model to describe it, which
corresponds to a 1D Bose gas going from a weakly to a strongly interacting
situation.

In order to solve the hydrodynamic equations
we have made use of a very recent approach which allows to find
analytical or quasi-analytical solutions of these equations
for a large class of models.
These model solutions allow to approximate very nearly the exact equation
of state which is the only input of the hydrodynamic equations coming
from the physical properties of the system. When in addition a first 
order correction has been made, we have been able to check that this
method gives the correct mode frequency with at least a relative precision of
$10 ^{-3}$ which is more than necessary for any practical purpose.
On the other hand the simplest of this model gives a very easy and convenient
analytical solution. Taken together the ensemble of these models allow to
cover all the range from simple analytical solutions to very precise
numerical solutions. We have compared our results to those obtained by
Menotti and Stringari from a sum rule approach, and we have found 
an excellent agreement. 

The method used in this
paper is quite powerful. It is not restricted to the lowest frequency mode 
and it 
can be used as well for any higher frequency mode. We have not
presented the corresponding results in the present paper only to avoid 
to make it oversized, but this would have been quite easy.
Another interest of our method is that it is not restricted to mean field
and can be applied to any equation
of state corresponding possibly to a very dense Bose gas. For example we
could very well consider the situation where the gas is dense enough so 
that the 1D intermediate density situation can no longer be described 
by mean field \cite{asgi,blume}.
Moreover the convenience
of analytical solutions makes it possible to invert the
method \cite{rcxl} and to extract the effective equation of state of the gas
from experimental data on the variation of the modes frequencies as
a function of the density.

We acknowledge numerous conversations with Y. Castin, C. Cohen-Tannoudji, 
J. Dalibard, D. Gangardt and C. Salomon.
We are extremely grateful to Sandro Stringari and Chiara Menotti for 
many stimulating discussions and for providing us with their data.

* Laboratoire associ\'e au Centre National de la Recherche Scientifique 
et aux Universit\'es Paris 6 et Paris 7.


\begin{references}
\bibitem{sign}We follow Menotti and Stringari
\cite{Menotti} in the definition of the
effective 1D scattering length. There is a difference in sign
with the definition of Olshanii \cite{Olshanii}.
\end{references}
\end{document}